\documentclass[aps,prl,reprint,balancelastpage,nofootinbib,preprintnumbers,superscriptaddress]{revtex4-1}

\usepackage{slashed}
\usepackage{bm}
\usepackage{latexsym,amssymb,amsmath,float,url,mathrsfs}
\usepackage{latexsym}
\usepackage{graphicx}
\usepackage{epstopdf}
\usepackage{amsmath}
\usepackage{comment}
\usepackage{subfigure}
\usepackage{natbib}
\usepackage{hyperref}
\usepackage{pifont}
\usepackage{blindtext}
\usepackage{adjustbox}
\usepackage{multirow}
\usepackage{tabularx}
\usepackage[table]{xcolor}
\usepackage{orcidlink}
\usepackage{tikz}
\usetikzlibrary{tikzmark}
\usepackage{fontawesome}

\usepackage{blkarray}
\usepackage{graphicx}
\usepackage{amsfonts}
\usepackage{soul}
\usepackage{amssymb}
\usepackage{amsmath}
\usepackage{cancel}
\usepackage{tcolorbox}

\usepackage{graphics,appendix,afterpage,makecell} 

\def\dd{{\rm d}}

\definecolor{oucrimsonred}{rgb}{0.6, 0.0, 0.0}
\definecolor{persianblue}{rgb}{0.11, 0.22, 0.73}
\definecolor{forestgreen}{rgb}{0.13,0.35,0.13}
\definecolor{lightgray}{rgb}{0.83, 0.83, 0.83}
 \hypersetup{colorlinks, citecolor=oucrimsonred, linkcolor=black, urlcolor=oucrimsonred}
\definecolor{cornellred}{rgb}{0.7, 0.11, 0.11}
\definecolor{navyblue}{rgb}{0.0, 0.0, 0.5}
\definecolor{amethyst}{rgb}{0.6, 0.4, 0.8}
\definecolor{yellow}{rgb}{1.0, 1.0, 0.0}
\definecolor{firebrick}{rgb}{0.7, 0.13, 0.13}
\definecolor{tangerineyellow}{rgb}{1.0, 0.8, 0.0}
\definecolor{deepfuchsia}{rgb}{0.76, 0.33, 0.76}
\definecolor{amber}{rgb}{1.0, 0.75, 0.0}
\definecolor{VioletRed4}{rgb}{0.55, 0.13, .32}
\definecolor{indiagreen}{rgb}{0.07, 0.53, 0.03}
\definecolor{VioletRed4}{rgb}{0.55, 0.13, .32}
\newcommand{\be}{\begin{equation}}
\newcommand{\ee}{\end{equation}}
\newcommand{\bea}{\begin{equation} \begin{aligned}}
\newcommand{\eea}{\end{aligned} \end{equation}}

\definecolor{oucrimsonred}{rgb}{0.6, 0.0, 0.0}
\newcommand\vertarrowbox[3][6ex]{%
  \begin{array}[t]{@{}c@{}} #2 \\
  \left\uparrow\vcenter{\hrule height #1}\right.\kern-\nulldelimiterspace\\
  \makebox[0pt]{\scriptsize#3}
  \end{array}%
}

\definecolor{verdechiaro}{rgb}{0.6,1,0.6}
\definecolor{giallochiaro}{rgb}{1,1,0.6}
\definecolor{bluscuro}{rgb}{0.15, 0.2, 0.9}
\definecolor{verdes}{rgb}{0.1, 0.5, 0.1}%
\definecolor{tangerineyellow}{rgb}{1.0, 0.8, 0.0}

\definecolor{americanrose}{rgb}{1.0, 0.01, 0.24}
\definecolor{cobalt}{rgb}{0.0, 0.28, 0.67}
\definecolor{brandeisblue}{rgb}{0.0, 0.44, 1.0}
\definecolor{mycolor}{rgb}{0.0, 0.0, 0.5}
\definecolor{oxfordblue}{rgb}{0.0, 0.13, 0.28}
\definecolor{azure}{rgb}{0.0, 0.5, 1.0}
\definecolor{turquoiseblue}{rgb}{0.0, 1.0, 0.94}
\newtcolorbox{mynewbox}[1]{colback=white!5!white,colframe=azure!75!black,fonttitle=\bfseries,title=#1}
\newtcolorbox{mybox}{colback=mycolor!5!white,colframe=azure!75!black}
\newtcolorbox{mynamedbox}[1]{colback=mycolor!5!white,colframe=azure!75!black,title=#1}
\definecolor{venetianred}{rgb}{0.78, 0.03, 0.08}
\newtcolorbox{mynamedbox1}[1]{colback=venetianred!5!white,colframe=venetianred!80!black,title=#1}
\newtcolorbox{mynamedbox2}[1]{colback=azure!5!white,colframe=azure!80!black,title=#1}

\definecolor{verdes}{rgb}{0.1, 0.5, 0.1}%
\definecolor{cornellred}{rgb}{0.7, 0.11, 0.11}

\definecolor{VioletRed4}{rgb}{0.55, 0.13, .32}

\hypersetup{
     colorlinks   = true,
     citecolor    = violet,
     urlcolor     = violet,
     linkcolor    = violet}

\definecolor{rossocorsa}{rgb}{0.83, 0.0, 0.0}

\usepackage[normalem]{ulem}

\newcommand{\papertitle}{The  AdS Perspective on  the Nonlinear Tails  in  Black Hole Ringdown}

\begin{document}

\title[]{\papertitle}

\author{Alex Kehagias\orcidlink{}}
\affiliation{Physics Division, National Technical University of Athens, Athens, 15780, Greece}
\affiliation{CERN, Theoretical Physics Department, Geneva, Switzerland}

\author{Antonio Riotto\orcidlink{0000-0001-6948-0856}}
\affiliation{Department of Theoretical Physics and Gravitational Wave Science Center,  \\
24 quai E. Ansermet, CH-1211 Geneva 4, Switzerland}


\begin{abstract}
\noindent
Black holes gradually settle into their static configuration by emitting gravitational waves, whose amplitude diminish over time according to a power-law decay at fixed spatial locations. We show that the nonlinear tails in the presence of a quadratic source,  which have been recently found  to  potentially dominate over the linear ones,   can be simply derived from  the  AdS$_2$$\times$S$^2$ spacetime perspective with  their amplitudes being  related to the Aretakis constants.

\end{abstract}

\maketitle

\noindent\textbf{Introduction --} 
The rise of gravitational wave (GW) astronomy has inaugurated a transformative period for performing high-precision tests of strong-field gravity \cite{Berti:2015itd,Berti:2018vdi,Franciolini:2018uyq,LIGOScientific:2020tif}. Ground-based interferometers -- LIGO, VIRGO, and KAGRA -- together with the upcoming space-based mission LISA, are attaining the sensitivities necessary to probe the ringdown phase of black hole (BH) mergers in detail \cite{Berti:2005ys,KAGRA:2013rdx,LIGOScientific:2016aoc,KAGRA:2021vkt,LIGOScientific:2023lpe}. A key component of this investigation involves characterizing the behavior of BH perturbations at late times, where both linear and nonlinear dynamics impart distinctive features onto the emitted GWs.

Historically, linear perturbation theory has formed the backbone of our understanding of BH ringdowns. This framework predicts that, at late times, BHs exhibit a characteristic pattern of exponentially decaying oscillations,  known as quasinormal modes (QNMs) (for a review, see Ref. \cite{Berti:2025hly}), followed by a late-time decay that adheres to an inverse power-law in time. Price’s well-known result exemplifies this decay: for massless fields propagating on a non-rotating BH background, the signal at fixed radius decays as $\sim t^{-2\ell-3}$ in the asymptotic regime, where $\ell$ denotes the multipole index \cite{Price:1971fb,Price:1972pw,Leaver:1986gd,Gundlach:1993tp,Gundlach:1993tn,Ching:1995tj,Chandrasekhar:1975zza,Martel:2005ir,Barack:1998bw,Berti:2009kk}.

To better capture the features of BH ringdowns, substantial work has gone into refining predictions for QNM spectra and late-time tails \cite{Ching:1994bd,Krivan:1996da,Krivan:1997hc,Krivan:1999wh,Burko:2002bt,Burko:2007ju,Hod:2009my,Burko:2010zj,Racz:2011qu,Zenginoglu:2012us,Burko:2013bra,Baibhav:2023clw}. These linear results have been strongly supported by numerical relativity and increasingly sensitive GW measurements. 

More recently, attention has shifted toward the nonlinear regime of BH perturbations \cite{Gleiser:1995gx,Gleiser:1998rw,Campanelli:1998jv,Garat:1999vr,Zlochower:2003yh,Brizuela:2006ne,Brizuela:2007zza,Nakano:2007cj,Brizuela:2009qd,Ripley:2020xby,Loutrel:2020wbw,Pazos:2010xf,Sberna:2021eui,Redondo-Yuste:2023seq,Mitman:2022qdl,Cheung:2022rbm,Ioka:2007ak,Kehagias:2023ctr,Khera:2023oyf,Bucciotti:2023ets,Spiers:2023cip,Ma:2024qcv,Zhu:2024rej,Redondo-Yuste:2023ipg,Bourg:2024jme,Lagos:2024ekd,Perrone:2023jzq,Kehagias:2024sgh,Bucciotti:2025rxa,bourg2025quadraticquasinormalmodesnull,BenAchour:2024skv,Kehagias:2025ntm}, uncovering new phenomena beyond linear predictions, such as second-order QNMs and their power-law tails.
The emergence of nonlinear tails holds significant interest, as they stem naturally from the outgoing QNM structures inherent in ringdown processes \cite{Okuzumi:2008ej,Lagos:2022otp}, reflecting the underlying nonlinear dynamics of gravity itself. Advances in numerical techniques have recently uncovered distinct power-law decay patterns in the Weyl scalar associated with these nonlinear features, setting them apart from traditional Price tails \cite{Cardoso:2024jme}. 
Furthermore, full 3+1-dimensional simulations of black hole mergers in numerical relativity have detected power-law decays that diverge from the classical Price tail expectations \cite{DeAmicis:2024eoy,Ma:2024hzq}. These developments enrich our understanding of nonlinear phenomena in black hole perturbation theory and underscore the importance of advancing experimental strategies—such as refined gravitational wave modeling—to probe their observational imprints.

Numerical analyses have also shown that a second-order $\ell=4$ mode, sourced by two interacting $\ell=2$ modes, exhibits a late-time decay proportional to $\sim t^{-10}$ \cite{Cardoso:2024jme} pointing toward a generalized behavior of the form $\sim t^{-2\ell-2}$. This nonlinear law
 has been shown to arise from the propagation of the quadratic QNMs in an asymptotically flat spacetime \cite{Ling:2025wfv,Kehagias:2025xzm}. 
In this paper, we take a further step ahead and show that such nonlinear tails may be understood from a symmetric
perspective once it is realized that the
spacetime in which the nonlinear tails
are generated is effectively an AdS$_2$$\times$S$^2$ spacetime. This allows to 
simply derive the nonlinear tail power-law from first principles and relate its amplitude to the so-called Aretakis constants in AdS$_2$.

\vskip 0.5cm
\vspace{-5pt}\noindent\textbf{The  Schwarzschild BH, the   AdS$_2$$\times$S$^2$ symmetry at infinity,  the ladder operators and the Green function --} 
Our starting point is a Schwarzschild BH with mass $M$ whose spacetime is described by the   metric 

\be
\label{metric}
{\rm d} s^2=-f(r){\rm d} t^2+\frac{{\rm d} r^2}{f(r)}+r^2{\rm d}\Omega_2^2, \,\,\,f(r)=1-r_s/r,
\ee
where $r_s=2G_NM$ is the Schwarzschild radius. 
In this  BH background, a massless scalar field $\Psi$  evolves according to $\Box \Psi(r,t)=0$, which after expanding $\Psi$ as 
\be
\Psi=\sum_{\ell=0}^\infty\sum_{m=-\ell}^{\ell}\frac{\psi_{\ell}(r,t)}{r} Y_{\ell m}(\theta,\phi)
\ee
where $Y_{\ell m}$ are scalar spherical 
harmonics, we find that $\psi_{\ell}(r,t)$
satisfies
\be
\left[\frac{\partial^2}{\partial r_*^2}
-\frac{\partial^2}{\partial t^2}-\left(1-\frac{r_s}{r}\right)\left( \frac{\ell(\ell+1)}{r^2}+\frac{r_s}{r}\right)\right]\psi_\ell=0.
\label{full}
\ee
As usual, $r_*$ is the tortoise coordinate defined as 
 $\dd r_*=\dd r/f$. At  large $r$, we have that $r=r_*$, and Eq. (\ref{full}) is written as 
 \be
 \left(\frac{\partial^2}{\partial r_*^2}
-\frac{\partial^2}{\partial t^2}- \frac{\ell(\ell+1)}{r_*^2}\right)\psi_\ell\approx 0,
\label{ap}
\ee
whereas the corresponding Green function satisfies 
\be
\left(\!\frac{\partial^2}{\partial r_*^2}
\!-\!\frac{\partial^2}{\partial t^2}\!-\! \frac{\ell(\ell\!+\!1)}{r_*^2}\!\right)G(t,r_*;t',r_*')\!=\!
\frac{\delta(t\!-\!t')\delta(r_*\!\!-\!r_*')}{\sqrt{-g}}.
\label{green0}
\ee
Let us now consider an exact  AdS$_2$$\times$S$^2$ spacetime with metric in Poincar\'e coordinates for the AdS$_2$ 
\begin{eqnarray}
\dd s^2&=&g_{ij}\dd x^i\dd x^j+\dd \Omega_2^2\nonumber \\
&=&\frac{-\dd t^2+\dd r_*^2}{r_*^2}+\dd \Omega_2^2, \qquad (i,j=0,1).
\label{ads}
\end{eqnarray}
A massless scalar field $\Psi$ on AdS$_2$$\times$S$^2$ obeys 
\be 
\Box\Psi=\Box_{\rm AdS_2} \Psi+\Box_{S^2}\Psi=0.
\label{box}
\ee
By using the metric (\ref{ads}) 
and expanding  $\Psi$ as $\Psi=\sum \psi_\ell(r_*,t) Y_{\ell m}(\theta,\phi)$ in spherical harmonics, we find that Eq. (\ref{box}) turns out to be  
\be
\Box_{\rm AdS_2} \psi_\ell-\ell(\ell+1)\psi_\ell=0.
\label{ads1}
\ee
Therefore, the modes $\psi_\ell$ are massive scalars on AdS$_2$, with square mass $m^2=\ell (\ell+1)$.
The scaling dimension of a field of mass $m$ in AdS$_2$ is 
\begin{equation}
    \Delta=\frac{1}{2}+\sqrt{\frac{1}{4}+m^2},
\end{equation}
 which for $m^2=\ell(\ell+1)$, gives 
\be
\label{weight}
\Delta=\ell+1.
\ee
Since on AdS$_2$ we have 
\be
\Box_{\rm AdS_2} =\frac{1}{\sqrt{-g}}\partial_i\left(\sqrt{-g}g^{ij}\partial_j\right)=r_*^2\left(\frac{\partial^2}{\partial r_*^2}
-\frac{\partial^2}{\partial t^2}\right),
\ee
Eq. (\ref{ads1}) is written as 
\be
r_*^2\left(\frac{\partial^2}{\partial r_*^2}
-\frac{\partial^2}{\partial t^2}\right)\psi_\ell-\ell (\ell+1)\psi_\ell=0,
\ee
which coincides with Eq. (\ref{ap}). In other words, the equation that dominates the dynamics at large $r$ of a massless scalar in the Schwarzschild BH background, is identical to the equation which governs a massive scalar with mass square $m^2=\ell(\ell+1)$ on an 
AdS$_2$ spacetime. 

Now, the Green function of Eq. (\ref{ads1}) satisfies \cite{Burgess:1984ti,Satoh:2002bc}
\be
\left(\Box_{\rm AdS_2} -m^2\right)G_{\rm AdS_2}=
\frac{1}{\sqrt{-g}}\delta(t-t')\delta(r_*-r_*').
\label{green}
\ee
The  solution to the above equation has been found at various occasions, and it is identical to the Green function of the BH Green function for large $r$ of Eq. (\ref{green0}).    For the conformal weight (\ref{weight}), it is  given by
\begin{align}
G_{\rm AdS_2}(\chi)=&\frac{1}{2^\ell\sqrt{\pi}}
\frac{\Gamma(\ell+1)}{\Gamma(\ell+\frac{3}{2})} 
\chi^{-\ell-1}\nonumber \\&{}_2F_1\left(\frac{\ell}{2}+\frac{1}{2},\frac{\ell}{2}+1,\ell+\frac{3}{2},\frac{1}{\chi^2}\right), 
\label{g1}
\end{align}
where $\chi$ is the AdS$_2$ invariant distance defined in the Appendix A. For $(t-t')\gg r_*>r_*'$ 
we find 
\be
\chi\approx -\frac{(t-t')^2}{2 r_* r_*'},
\ee
so that Eq. (\ref{g1}) gives 
\be
G_{\rm AdS_2}(t,r_*;t',r_*')\approx 
-(-1)^\ell \frac{2}{\sqrt{\pi}}
\frac{\Gamma(\ell+1)}{\Gamma(\ell+\frac{3}{2})} \, \frac{(r_*r_*')^{\ell+1}}{(t-t')^{2\ell+2}}.
\label{g2}
\ee
Note that the Green function in Eq. (\ref{g1}) is invariant under the AdS$_2$ isometries, namely dilations, time translations and conformal inversions. One may then wonder where such symmetries are in the standard determination of the tails from the Green function branch cut. The latter is given by (setting $t'=0$)
\cite{Andersson:1996cm}
\be
G^C(t,r;r')=4iM r r'\int_0^{-i \infty} \dd \omega\,\omega^2 j_\ell(\omega r)j_\ell(\omega r').
\ee
Evaluating the above integral leads to 
\be
G^C(t,r;r')=-\frac{2M}{\pi}\frac{t}{r r'}\frac{Q_\ell(\chi)}{\sqrt{\chi^2-1}}, 
\label{g3}
\ee
where $Q_\ell(\chi)$ is the Legendre polynomial. We have been able  the latter in terms of hypergeometric function,  
Eq. (\ref{g3}) can be written as 
\begin{align}
G^C(t,r;r')=&-\frac{M}{2^\ell\sqrt{\pi}}\frac{t}{r r'}\frac{\Gamma(\ell+1)}{\Gamma(l+\frac{3}{2})}\frac{\chi^{-\ell-1}}{\sqrt{\chi^2-1}}\nonumber \\
&{}_2F_1\left(\frac{\ell}{2}+\frac{1}{2},\frac{\ell}{2}+1,\ell+\frac{3}{2},\frac{1}{\chi^2}\right).
\label{g4}
\end{align}
Comparing Eqs. (\ref{g4}) and (\ref{g1}) at $r\simeq r_*$, we see that 
\be
G^C(t,r;r')=\frac{M t}{rr'} \frac{G(\chi)}{\sqrt{\chi^2-1}}.
\ee
In the limit $t\gg r$, or $\chi^2\gg 1$, we find 

\begin{eqnarray}
G^C(t,r_*;r_*')&=&\frac{2M}{t}G_{\rm AdS_2}(\chi)\sim \frac{1}{t^{2\ell+3}}\ll G_{\rm AdS_2}(\chi)
\nonumber\\
&\sim &\frac{1}{t^{2\ell+2}},
\end{eqnarray}
and we see that the contribution to the Green function from the branch cut is subleading with respect to the flat spacetime contribution when $t\gg r$.

Let us return now to the Green function in Eq. (\ref{g1}). The latter is actually the two-point function calculated in the 
SO(2,1)-invariant vacuum in Poincar\'e coordinates $|0_{\rm Poincare}\rangle$ \cite{Spradlin:1999bn}. 
Then, by using the standard relation
between Legendre and hypergeometric functions
\begin{align}
 Q_\ell(z)=&2^{-\ell-1} \frac{\Gamma{(\ell+1)}}{\Gamma(\ell+\frac{3}{2})}z^{-\ell-1}\nonumber \\
 &\cdot {}_2F_1\left(1+\frac{\ell}{2}+\frac{1}{2}+\frac{\ell}{2},\ell+\frac{3}{2},\frac{1}{z^2}\right),  |z|>1, \label{QF} \end{align} 
we can express Eq. (\ref{g1}) as 
\begin{eqnarray}
G_{\rm AdS_2}(\chi)=G^\ell_{\rm AdS_2}(\chi)=\frac{1}{2\pi}Q_\ell(\chi). \label{gQ}
\end{eqnarray}
Using the recurrence relation of the Legendre polynomials, we can define raising and lowering ladder operators $L_\pm$ as 
\begin{align}
    L_+=&(\chi^2-1)\frac{\rm d}{{\rm d}\chi}+(\ell+1)\chi, \nonumber \\
     L_-=&(1-\chi^2)\frac{\rm d}{{\rm d}\chi}+\ell\chi, \label{ladder}
\end{align}
which satisfy
\begin{align}
    L_+G^\ell_{\rm AdS_2}(\chi)=&(\ell+1)G^{\ell+1}_{\rm AdS_2}(\chi),
    \nonumber \\
    L_-G^\ell_{\rm AdS_2}(\chi)=&\ell\,G^{\ell-1}_{\rm AdS_2}(\chi). 
\end{align}
In addition, if we further define the operator $L_0$ such that 
\be
L_0\,G^\ell_{\rm AdS_2}(\chi)=\left(\ell+\frac{1}{2}\right)G^\ell_{\rm AdS_2}(\chi),
\ee
then it is easy to verify that $L_\pm$ and $L_0$ satisfy the SO$(2,1)$ algebra 
\be
[L_0,L_\pm]=\pm L_\pm, \qquad
[L_+,L_-]=-2 L_0.
\ee
Therefore, starting with the lowest $\ell=0$ Green function $G^0_{\rm AdS_2}(\chi)$, we can construct all $G^\ell_{\rm AdS_2}(\chi)$'s by repeated applications of the raising operator $L_+$ as 
\be
G^\ell_{\rm AdS_2}(\chi)=\frac{1}{\ell!}L_+^\ell G^{\ell=0}_{\rm AdS_2}(\chi).
\ee
In addition, let us note that the  SO$(2,1)$ spectrum generating symmetry mentioned above is related to the hidden AdS$_2$  structure underlying our problem. 

To proceed, let us notice that $Q_\ell(z)$ is analytic in the whole complex z-plane with branch points at $z=\pm 1, \infty$, whereas the standard branch cut is along the $(-\infty,1]$. Actually, in the complex $z$ plane $Q_\ell(z)$ is defined by Eq. (\ref{QF}). The hypergeometric function  has a branch cut from $z=1$ to $z=\infty$ along the real axis, and defining  
$z^{-\ell-1}=e^{-(\ell+1)\log z}$, the branch cut is taken to be $(-\infty,1]$. Then, the Feynman propagator is the limit of $G_{\rm AdS_2}(\chi+i0)$ above the cut, whereas the retarded Green function is 
\begin{eqnarray}
G^{\rm ret}_{\rm AdS_2}(\chi)&=&i\theta(t-t') 
\Big[G_{\rm AdS_2}(\chi+i 0)-G_{\rm AdS_2}(\chi-i 0)\Big]. \nonumber\\
&&
\label{gr1}
\end{eqnarray}
Then by using the relation \cite{abramowitz_64}
\begin{eqnarray}
Q_\ell(z+i0)=Q_\ell(z-i0)-i\pi P_L(\chi), 
\end{eqnarray}
we find that the retarded Green function is given by
\begin{equation}
G^{\rm ret}_{\rm AdS_2}(\chi)=\frac{1}{2}\theta(t-t')P_L(\chi).
\label{Gret} 
\end{equation}

\begin{figure}[t!]
\centering
  \includegraphics[width=0.50\textwidth]{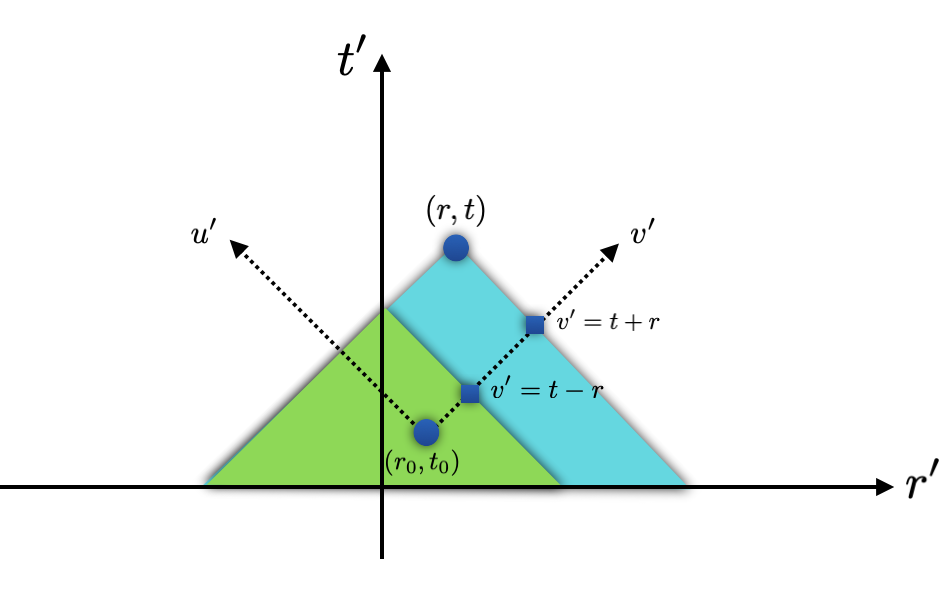}
  \caption{The causality diagram concerning an outgoing signal propagating from the time $t_0$, emitted at the point $r_0$, at the speed of light along the null direction $u'=(t_0-r_0)=u_0$.}
  \label{fig:causality}
\end{figure}

\vspace{5pt}\noindent\textbf{The nonlinear quadratic source  --} 
Imagine now that there is a second-order source composed of the combination of two linear QNMs (e.g. an $L=4$ mode made out of two linear $\ell=2$ modes) and decaying as $\sim 1/r^2$ at infinity \cite{Nakano:2007cj}. We assume that such a source is  initiated at a specific moment $t_0$ and radial coordinate $r_0$  and then propagates outward at the speed of light. 
Its influence is confined in retarded time, denoted by $u' \equiv (t' - r') \simeq u_0\pm \delta$, where $\delta$ indicate the small support of the propagating source. In contrast, the advanced time variable $v' \equiv (t' + r')$ spans the range from $(t-r)$ to $(t+r)$, see Fig. 1. We also assume that the observer's position is such that the time coordinate is significantly larger than the radial  coordinate, i.e., $t \gg r$. Under these circumstances \cite{Okuzumi:2008ej,Ling:2025wfv,Kehagias:2025xzm}, and from the the AdS$_2$$\times$S$^2$ perspective, the  nonlinear quadratic $L$-mode satisfies an equation of the form 

\be
\left(\Box_{\rm AdS_2} -L(L+1)\right)\psi_L=A_L\, e^{-2i\omega_\ell u}.
\label{green}
\ee
The nonlinear quadratic tail
is therefore fully dictated by the behavior of the AdS$_2$ Green function 

\begin{align}
\psi_L&=&\int_{u_0-\delta}^{u_0+\delta}{\rm d}u'\,e^{-2i\omega_\ell u'}
\int_{t-r}^{t+r}\frac{1}{2} P_L(\chi) \left(1-\frac{2 M}{r'}\right)^{-1} {\rm d} v',\nonumber\\ 
\label{in1}
\end{align}
where  we have gone back to the variable $r$, and we write the second integral as 

\begin{align}
I_2(u') =&\int_{t-r}^{t+r}\frac{1}{2} P_L(\chi) \left(1-\frac{2 M}{r'}\right)^{-1} {\rm d} v' \nonumber \\
=&\frac{1}{2}\sum_{n=0}^\infty \int_{t-r}^{t+r} P_L(\chi)\left(\frac{2M}{r'}\right)^n  {\rm d} v'. \label{in2}
\end{align}

In terms of $u', v'$, $\chi$ is written as 
\begin{eqnarray}
\chi=\frac{r^2-(t-u')(t-v')}{t(u'-v')}
\end{eqnarray}
and if we trade   $v'$ for $\chi$ in the integral of Eq. (\ref{in2}), we get 
\begin{align}
 I_2\!=&\sum_{n=0}^\infty  \!r \Big[r^2\!-\!(t\!-\!u')^2\Big]^{-1-n}\left(2M\right)^{n}\nonumber \\
 &\int_{-1}^1 \!\! {\rm d} \chi P_L(\chi) (r\chi\!+\!t\!-\!u')^n.
  \label{in3}
 \end{align} 
Then, using the relation
\begin{eqnarray}
\int_{-1}^1 {\rm d} \chi P_L(\chi)(a\chi+b)^n=0, \qquad\mbox{for}
\qquad n<L,
\end{eqnarray}
and
\begin{eqnarray}
\int_{-1}^1P_L(\chi)(a\chi+b)^L{\rm d} \chi=\frac{2^L L!(L-1)!}{(2L+1) (2L-1)!!}a^L,
\end{eqnarray}
we find the first non zero contribution appears for $n=L$ for which
\begin{align}
 I_2\!=&r\!\Big[r^2\!-\!(t\!-\!u')^2\Big]^{-1-L}\!\!(2M)^{L}\!\!\int_{-1}^1 P_L(\chi) (r\chi+t-u')^L {\rm d} \chi\nonumber\\
  =& \dfrac{2^L L!(L-1)!}{(2L+1) (2L-1)!!}\dfrac{(2M)^{L}\,r^{L+1}}{\Big[r^2-(t-u')^2\Big]^{1+L}}.  \label{in4}
 \end{align} 
For $t\gg r$ 
we find 
\be
I_2\approx (-)^{L+1}\dfrac{ 2^L L!(L-1)!(2M)^{L}}{(2L+1) (2L-1)!! }\, \frac{r^{L+1}}{t^{2L+2}},
\label{I2}
\ee
which reproduces the findings of Refs. \cite{Ling:2025wfv,Kehagias:2025xzm} as the remaining integral in $u'$ amounts to replacing $u'$ with $u_0$ for $u_0\ll \delta$. The generation mechanism for the nonlinear quadratic tail differs from that of the linear Price's tail. The latter is sourced by QNMs back-scattered from the potential at infinity; this potential peaks at small, nonvanishing frequencies, resulting in an amplitude proportional to the BH mass. Conversely, the nonlinear tail originates from the quadratic QNM itself. Because this mode is constant along the edge $(t'-r_*')\simeq u_0$, it does not show damped oscillations. 

\vspace{5pt}\noindent\textbf{Relating the nonlinear quadratic ringdown tails to the Aretakis constants --} 
Let us now show the relation of our result with the so-called Aretakis constants \cite{Aretakis:2011ha, Aretakis:2011hc,Aretakis:2012ei}. First, we
 define new coordinates
\be 
R=\frac{1}{r}, \qquad v=t-\frac{1}{R},
\ee
so that the AdS$_2$ part of the metric in Eq. (\ref{ads}) is written as 
\begin{eqnarray}
\dd s^2=-R^2\dd t^2+\frac{\dd R^2}{R^2}=-R^2 \dd v^2+2 \dd v \dd r.
\label{ads2}
\end{eqnarray}
Then, the Aretakis number $\mathcal{H}_L$ \cite{Aretakis:2011hc,Aretakis:2011ha} for a field in AdS$_2$  of squared mass $L(L+1)$ (more details  in Appendix B) reads
\be
\mathcal{H}_L=
\left(\partial_R^{L+1} \psi\right)\Big|_{R=0},
\ee
and is constant 
\be
\partial_v \mathcal{H}_L=0.
\ee
In particular, for the scalar field 
$\psi_L$, the Aretakis number is expressed as \cite{Lucietti:2012xr} 

\begin{align}
\mathcal{H}_L=&
\left(\partial_R^{L+1} \psi_L\right)\Big|_{R=0}
=
\left(\partial_R^{L+1} I_1 I_2\right)\Big|_{R=0}.
\end{align}
Since in the  new coordinates $(R,v)$, the integral $I_2$ in Eq. (\ref{I2}) is written as 
\be
I_2\approx -\dfrac{(-)^L 2^L L!(L-1)!}{(2L+1) (2L-1)!!}\frac{(2M)^{L}R^{L+1}}{(R v-1)^{2L+2}},
\label{I22}
\ee
we find that 
\be
\left(\partial_R^{L+1}
I_2\right)\Big|_{R=0}= (-)^{L+1}\dfrac{ 2^L \,(L!)^2\,(L-1)!(2M)^L}{(2L+1) (2L-1)!!}, 
\ee
so that 
\be
\mathcal{H}_L= -(-)^{L}\dfrac{ 2^L \,(L!)^2\,(L-1)!(2M)^L}{(2L+1) (2L-1)!!}I_1. 
\ee
Then, in the original $(t,r)$ coordinates, we can write  
\be
I_2\approx \frac{L!}{I_1} \mathcal{H}_L \frac{r^{L+1}}{t^{2L+2}}.
\ee
Therefore
\be
\psi_L\approx
L! \, \mathcal{H}_L \, 
\frac{r^{L+1}}{t^{2L+2}},
\ee
and the nonlinear source induces a tail proportional to the Aretakis constant.

\vspace{5pt}\noindent\textbf{Conclusions --} 
Nonlinear tails of the QNMs are getting recently attention due to the fact that
their power-law is smaller that the one dictated by Price's law at the linear level and therefore they might dominate
at at sufficiently long times. In this paper we have shown that, at the quadratic level, the dynamics of the tails can be understood from an AdS perspective. The question is if this conclusion holds at any order in perturbation theory  as higher-order  contributions, even though with  smaller amplitudes,   might, in principle, alter the power-law. We intend to address  this issue in  the next future.

\vskip 0.1cm
\centerline{\it Acknowledgments}
  \vskip 0.1cm
  \noindent
We thank S. Aretakis for discussions.   A.R.  acknowledges support from the  Swiss National Science Foundation (project number CRSII5\_213497) and from  the Boninchi Foundation through  the project ``PBHs in the Era of GW Astronomy''.

\appendix



\setcounter{equation}{0}
\renewcommand{\theequation}{A.\arabic{equation}}

\vspace{0.2cm}
\noindent
\section{Appendix A: The 
A\MakeLowercase{d}S$_2$$\times$S$^2$ invariant distance}
\noindent
\label{supp:1}
The AdS$_2$ spacetime (of unit radius) is the hyperboloid 
\be
X_0^2-X_1^2+X_2^2=1,
\label{hype}
\ee
embedded into 3D spacetime with metric
\be
\dd s^2=\eta_{MN} \dd X^M\dd X^N=-\dd X_0^2+\dd X_1^2-\dd X_2^2,
\ee
so that $\eta_{MN}=(-1,1,-1)$, $(M,N=0,1,2)$.
The parametrization of the AdS$_2$ hyperboloid (\ref{hype}) by the Poincar\'e coordinates $(t,r)$ 
\begin{align}
    X_0=&\frac{1}{2r}\left(r^2-t^2+1\right), \nonumber \\
    X_1=&\frac{1}{2r}\left(r^2-t^2-1\right), \nonumber \\
    X_2=&\frac{t}{r}, 
    \label{poin}
\end{align}
leads to the induced metric on the 
AdS$_2$
\be
\dd s^2=\frac{-\dd t^2+\dd r^2}{r^2}. \label{ss}
\ee
The invariant distance on the AdS$_2$ hyperboloid (\ref{hype}) between the points $(X_0,X_1,X_2)$
and $(X_0',X_1',X_2')$ is 
\be
\sigma(X,X')=\frac{1}{2}\eta_{MN}
(X-X')^M(X-X')^N,
\ee
which, by using the parametrization 
(\ref{poin}) is written as 
\be
\sigma=\frac{-(t-t')^2+(r-r')^2}{2 r r'}.
\label{s}
\ee
One quantity that appears frequently is 
\be
\chi=1+\sigma=\frac{-(t-t')^2+r^2+{r'}^2}{2 r r'}.
\label{x}
\ee
Both $\sigma$ and $\chi$ are invariant under the SL$(2)$ isometry group of AdS$_2$ corresponding to the invariance of the metric (\ref{ss}) under dilations $x^i\to \lambda x^i$, time shift $t\to t+\mbox{constant}$, and conformal inversions $x^i\to x^i/x_j x^j$.

\section{Appendix B: Aretakis number}
\setcounter{equation}{0}
\renewcommand{\theequation}{B.\arabic{equation}}
\noindent
\label{supp:1}
\noindent
Consider Eq. (\ref{ap}) which we write here as
\be
 \left(\frac{\partial^2}{\partial r^2}
-\frac{\partial^2}{\partial t^2}- \frac{\ell(\ell+1)}{r^2}\right)\psi_\ell= 0.
\label{appp}
\ee
One can recognize the above equation as the KG in AdS$_2$  with mass square $\ell (\ell+1)$. 
Such  mass is the result of a massless scale on AdS$_2\times$S$_2$. Indeed, 
the AdS$_2\times$S$_2$ spacetime has the metric

\begin{eqnarray}
\dd s^2&=&g_{ij}\dd x^i\dd x^j+\dd \Omega_2^2\nonumber \\
&=&\frac{-\dd t^2+\dd r^2}{r^2}+\dd \Omega_2^2, \quad (i,j=0,1).
\end{eqnarray}
for which the  AdS$_2$ part has inverse  $g^{ij}={\rm diag}(-r^2,r^2)$ and $\sqrt{-g}=1/r^2$. 
In  AdS$_2$ we have 
\begin{align}
\Box_{\rm AdS_2} =&\frac{1}{\sqrt{-g}}\partial_i\left(\sqrt{-g}g^{ij}\partial_j\right)\nonumber \\
=&r^2\left(\frac{\partial^2}{\partial r^2}
-\frac{\partial^2}{\partial t^2}\right),
\end{align}
Eq. (\ref{appp}) is written as 
\be
r^2\left(\frac{\partial^2}{\partial r^2}
-\frac{\partial^2}{\partial t^2}\right)\psi_\ell-\ell (\ell+1)\psi_\ell=0,
\ee
In Poincar\'e coordinates, the AdS$_2$ metric can be written as
\begin{eqnarray}
\dd s^2&=&
-R^2\dd T^2+\frac{\dd R^2}{R^2},
\end{eqnarray}
for which $g^{ij}={\rm diag}(-1/R^2,R^2)$ and $\sqrt{-g}=1$.
In these coordinates we have 
\begin{align}
\Box_{\rm AdS_2} =&\frac{1}{\sqrt{-g}}\partial_i\left(\sqrt{-g}g^{ij}\partial_j\right)=\partial_R(R^2\partial_R)-\frac{1}{R^2}\partial_T^2\nonumber \\
=&\frac{1}{R^2}\left[R^2\partial_R(R^2\partial_R)-\partial_T^2\right].
\end{align}
Setting $R=1/r$ and $T=t$, one has 

\be
{\rm d}R=-\frac{{\rm d}r}{r^2} \,\,\,{\rm and}\,\,\, R^2\partial_R=-\partial_r
\ee
and

\be
\frac{1}{R^2}\left[R^2\partial_R(R^2\partial_R)-\partial_T^2\right]=r^2\left(\partial_r^2
-\partial_t^2\right).
\ee
We define 

\be
u=T+\frac{1}{R}=t+r\,\,\,{\rm and}\,\,\,
v=T-\frac{1}{R}=t-r,
\ee
for which ${\rm d}T={\rm d}v-{\rm d}R/R^2$ and the 
 metric becomes

\begin{align}
\dd s^2=&
-R^2\dd T^2+\frac{\dd R^2}{R^2}=-R^2\left(\dd v^2+\frac{\dd R^2}{R^4}-2\dd v\frac{\dd R}{R^2}\right)\nonumber\\
&+\frac{\dd R^2}{R^2}
=-R^2\dd v^2+2\dd v\dd R.
\end{align}
The equation for a massive scalar field of mass square $m^2=\ell (\ell+1)$ becomes

\be
2\partial_v\partial_{R}\psi_\ell+\partial_{R}(R^2\partial_{R}\psi_\ell)-\ell(\ell+1)\psi_\ell=0.
\label{msc}
\ee
Operating both sides with $\partial_{R}^\ell$, one encounters

\be
\partial_{R}^{\ell+1}(R^2\partial_{R}\psi_\ell)=\sum_{k=0}^{\ell+1}\left(\!\!\begin{array}{c}\ell+1\\ k\end{array}\!\!\right)\partial_{R}^{\ell +1-k}(R^2)\partial_{R}^{k+1}\psi_\ell.
\ee
For $R\simeq 0$, the sum does not vanish only for $(\ell+1-k)=2$, and one has 

\begin{align}
\partial_{R}^{\ell+1}(R^2\partial_{R}\psi_\ell)\simeq&2\left(\!\!\begin{array}{c}\ell\!+\!1\\ \ell\!-\!1\end{array}\!\!\right)\partial_{R}^{\ell}\psi_\ell\nonumber \\
=&\ell(\ell+1)\partial_{R}^{\ell}\psi_\ell,
\end{align}
from which we deduce the Aretakis constant \cite{Aretakis:2012ei}

\be
\partial_v\partial_{R}^{\ell+1}\psi_\ell\Big|_{R=0}=0\Rightarrow {\cal H}_\ell\equiv \partial_{R}^{\ell+1}\psi_\ell\Big|_{R=0}={\rm constant}.
\ee

\bibliography{main}

\end{document}